\documentclass[prl, twocolumn]{revtex4-1} 
\usepackage{amsmath}  
\usepackage{amsfonts} 
\usepackage{graphicx} 
\usepackage{subfigure}

\begin{document}
\bibliographystyle{unsrt}
\title{A testable prediction from entropic gravity}
\author{Mehmet S\"uzen}
\date{\today}
\begin{abstract}

I have shown conceptually that quantum state has a direct relationship to gravitational constant due to
entropic force posed by Verlinde's argument and part of the Newton-Schr\"odinger equation (N-S) 
in the context of gravity induced collapse of the wavefunction via Di\'osi-Penrose proposal. 
This direct relationship can be used to measure gravitational constant using state-of-the-art 
mater-wave interforemetry to test the entropic gravity argument.
\end{abstract}

\maketitle


\section{Introduction}

Recently, Verlinde has put forward an argument regarding entropic force induced
gravity \cite{verlinde11a}. On the other hand, Newton-Schr\"odinger equation is developed
to couple gravity with the collapse of a wavefunction, Penrose-Di\'osi \cite{diosi87a, diosi89a, penrose96a, diosi07a}.
Wavefunction collapse implies that when quantity of mass approaching to macroscopic scales, 
the wavefunction tends to localize in position \cite{vanmeter11a}. Here we aim at using these
two approaches to drive a consistent relationship between quantum state of atomic system
and the gravitational constant.

\section{Entropic Wavefunction} 

Unruh radiation is observed for an accelerated particle with acceleration $a$ \cite{unruh76a} with
temperature $T$,
\begin{eqnarray}
 \label{eq:unruh}
 T & = & \frac{1}{2\pi} \frac{\hbar a }{k_{B}c}
\end{eqnarray}

where, $\hbar$ is the Planck constant, $k_{B}$ is Boltzmann constant and $c$ is the 
speed of light. If we consider this particle placed in front of a membrane, 
holographic screen \cite{verlinde11a}, its distance and entropy associated with the
information on the membrane will read as follows,
\begin{eqnarray}
 \label{eq:dx}
 \Delta x=\frac{\hbar}{mc},
\end{eqnarray}
\begin{eqnarray}
 \label{eq:s}
 \Delta S=2\pi k_{B}.
\end{eqnarray}
This originates from Bekenstein's thought experiment \cite{verlinde11a}. The force $F$ excerted by the 
membrane on the particle reads,
\begin{eqnarray}
 \label{eq:f0}
  F & = & T \frac{\Delta S}{ \Delta x}.
\end{eqnarray}
We can plug Eqs. (\ref{eq:unruh}), (\ref{eq:dx}) and (\ref{eq:s}) into Eq. (\ref{eq:f0}).
This will lead to Newton's Law, with scalar field, particle's own graviational field, $\phi(x)$,
\begin{eqnarray}
 \label{eq:f1}
 F  = m a = - \nabla \phi(x) = - \frac{d \phi(x)}{dx}.
\end{eqnarray}

If we consider particle as a quantum mechanical object, moving in its own gravitational field,
the Schr\"odinger-Newton equation (S-N), are the pair of coupled nonlinear partial differential
equations \cite{harrison03a}, 

\begin{eqnarray}
 \label{eq:sn0}
 \nabla^{2} \phi(x)                                  & = & 4 \pi G m |\psi|^{2}, \\
  - \frac{\hbar^2}{2m} \nabla^{2} \psi - m \phi \psi & = & \tilde{E} \psi.
\end{eqnarray}

where $\phi(x)$ is the scalar field, $m$ is the mass, $\psi$ is the wavefunction,
$G$ is the gravitational constant and $\tilde{E}$ is the energy. Argument due to 
Penrose \cite{penrose96a} is that multiple quantum states reduce to one of the 
states in finite time because of the significant mass displacement.

Re-writing the first part of S-N with the entropic force, Eq.(\ref{eq:sn0}) 
with Eqs.(\ref{eq:f0}) and (\ref{eq:f1}) in one dimension,
\begin{eqnarray}
  \frac{d}{dx}(F)                            & = & 4 \pi G m |\psi|^{2}, \\
  \frac{d}{dx}(-\frac{d\phi(x)}{dx})         & = & 4 \pi G m |\psi|^{2}, \\
  \label{eq:f2}
 -\frac{d}{dx} (T \frac{\Delta S}{\Delta x}) & = & 4 \pi G m |\psi|^{2}.
\end{eqnarray}
Integrating both sides on Eq.(\ref{eq:f2}) and recall $\Delta x$
\begin{eqnarray}
  \label{eq:s1}
  -T \frac{m c \Delta S }{ \hbar} + C= & 4 \pi G m \int |\psi|^{2} dx.
\end{eqnarray},
If we write the entropy with von Neumann entropy {\it without the trace for collapsed state},
\begin{eqnarray}
 \label{eq:s2}
 \Delta S = - k_{B} \hat{\rho} \ln{\hat{\rho}},
\end{eqnarray}
where $\hat{\rho}$ is the density matrix, it's proportional to collapsed density,
\begin{eqnarray}
 \label{eq:sim}
 \hat{\rho} \sim \int |\psi|^{2} dx.
\end{eqnarray}
Using Eq.(\ref{eq:s2}) and (\ref{eq:sim}) into Eq.(\ref{eq:s1}), with $T$ and $\Delta x$,
\begin{eqnarray}
  \frac{-1}{2 \pi} \frac{\hbar a}{k_{B} c} \frac{m c k_{B} ln \hat{\rho}}{\hbar} & \sim & 4 \pi G m \hat{\rho}, \\
 \label{eq:rel}
 \hat{\rho} & \sim &  exp(-8\pi^{2} G/a). 
\end{eqnarray}

\section{Conclusion}
The proposed relationship at Eq. (\ref{eq:rel}) could be used in measuring $G$ using atom interforemeter \cite{fixler07a}
to varify entropic gravity argument.  However, mass of the atomic system should be $10^{9} u$ 
to go into S-N collapse \cite{vanmeter11a}, this might be an experimental challange \cite{vanmeter11a}, 
and density matrix of the atomic system should be computable from the first-principles.

\bibliography{lib}
\end{document}